\documentclass[prd,aps,showpacs,reprint,preprintnumbers,amsmath,amssymb,superscriptaddress,longbibliography,nofootinbib,floatfix]{revtex4-2}

\usepackage[utf8]{inputenc}
\usepackage{amsmath}
\usepackage{amssymb} 
\usepackage{enumerate}
\usepackage{multirow}
\usepackage{braket}
\usepackage[toc,page]{appendix}
\usepackage{url}
\usepackage{graphicx}
\usepackage[pdftex,bookmarks,linktocpage,pdfpagelabels,plainpages=false,hyperfigures,linkcolor=blue,citecolor=blue]{hyperref} 
\hypersetup{colorlinks=true}
\usepackage{adjustbox}
\usepackage{mathrsfs,amssymb}  
\usepackage{cancel}
\usepackage[normalem]{ulem}
\usepackage{cleveref}
\usepackage{tikz}
\usetikzlibrary{decorations.markings}
\usetikzlibrary{positioning}
\usetikzlibrary{shapes}
\usetikzlibrary{calc}

\newcommand{\beq}{\begin{equation}}
\newcommand{\eeq}{\end{equation}}

\newcommand{\bea}{\begin{eqnarray}}
\newcommand{\eea}{\end{eqnarray}}

\begin{document}

\title{Transformer networks for Heavy flavor jet tagging } 

\author{A.\,Hammad}
\email{hamed@post.kek.jp}
\affiliation{Theory Center, IPNS, KEK, 1-1 Oho, Tsukuba, Ibaraki 305-0801, Japan.}

\author{Mihoko M. \,Nojiri}
\email{nojiri@post.kek.jp }
\affiliation{Theory Center, IPNS, KEK, 1-1 Oho, Tsukuba, Ibaraki 305-0801, Japan.}
\affiliation{The Graduate University of Advanced Studies (Sokendai), 1-1 Oho, Tsukuba, Ibaraki 305-0801, Japan.}
\affiliation{Kavli IPMU (WPI), University of Tokyo, 5-1-5 Kashiwanoha, Kashiwa, Chiba 277-8583, Japan.}
\date{\today}
\begin{abstract}
In this article, we review recent machine learning methods used in challenging particle identification of heavy-boosted particles at high-energy colliders.   
Our primary focus is on attention-based Transformer networks. We report the performance of state-of-the-art deep learning networks and further improvement coming from the modification of networks based on physics insights.  Additionally, we discuss interpretable methods to understand network decision-making, which are crucial when employing highly complex and deep networks.
 \end{abstract}
 
\maketitle

\section{\bf Introduction}  
\noindent
The success of modern physics stems from the discovery of the fundamental principles of nature. Starting from the discovery of quantum mechanics and relativity, our understanding keeps expanding various boundaries toward the understanding of the origin of spacetime and matters in our Universe or complex physics systems.  Neither theory nor experiment only can explore the mystery of our universe. Achievement of modern physics comes from interplay between theory-experiment-technologies.  

The current highest energy particle collider is the Large Hadron Coliider (LHC) at CERN, which smashes the proton, creating heavy particles responsible for the fundamental interaction. Among those, the Higgs boson is responsible for the masses of all elementary particles in the standard model, and its nature has only recently started being explored. Higgs boson is the only scalar particle among the Standard Model (SM) particles, and its vacuum expectation value breaks the gauge symmetry of the SM spontaneously. Assuming that the SM describes the particle contents and interaction in nature, the measured Higgs boson and top quark masses suggest that the vacuum is metastable. To realize the current universe from the violent initial state condition of our universe, the SM Higgs boson interactions should be modified so that the current vacuum is absolutely stable. The precise measurements of the Higgs boson and the particles that couple strongly to the Higgs boson are important to find out the deviation from the SM assumptions. As the deviation is expected to be larger for high energy interactions, the identification of the Lorentz-boosted Higgs boson, top quark (the heaviest quark) and gauge bosons have been studied intensively. 

Identification of boosted particles is highly challenging. Higgs boson produced at the collider immediately decays into lighter particles (quarks or leptons). If it decays into quarks, the quarks further undergo the parton shower process, creating many quarks and gluons (partons) going in the same direction. Because of the strong interactions, such quarks and gluons create bound states, ending up a collimated spray of particles originating from a high-energy collision; the object is called a ``jet". The production of jets from other QCD processes overwhelms the $H\rightarrow bb$ process.  Identifying boosted heavy particles while rejecting quark and gluon background is called the jet identification problem. 

The task of the jet identification in hadron collider experiments has undergone significant development over the past decades. In particular, machine learning (ML) has revolutionized how we analyze and leverage various observables associated with jets and their constituents. 
While viewing the jet as a single entity provides limited information about the initiating particle, analyzing its constituents offers additional insights for predicting the initiating particle. Over the years, various approaches collectively known as jet tagging have been developed for this purpose.

\begin{figure}
    \centering
    \includegraphics[width=0.8\linewidth]{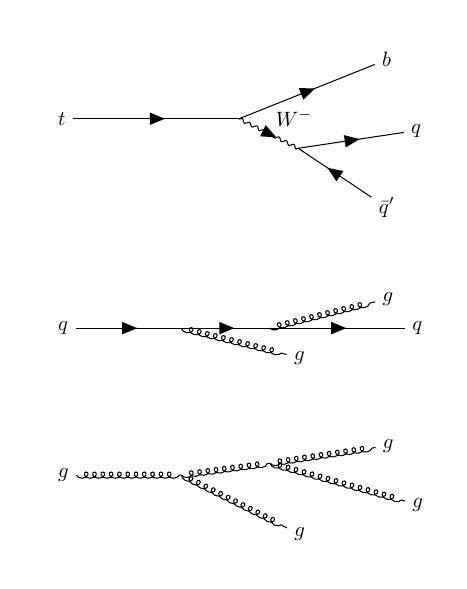}
    \caption{schematical figure of top quark decay into b and W  followed by W boson decay (top) and quark and gluon parton shower process (bottom two). }
    \label{fig:0}
\end{figure}
In the quest to analyze collision data and identify interesting physics processes, it is necessary to select a jet reconstruction strategy and a specific jet tagging algorithm based on the expected features of the final state. For instance, if the process involves producing a hadronically-decaying heavy particle (such as a W, Z, or Higgs boson, or a top quark) with a large transverse momentum, a large-radius "fat" jet may be used to reconstruct the entire decay within the jet radius. It was first realized in Ref. \cite{Butterworth:2008iy} that such jets are often characterized by a multi-pronged structure, and rejecting QCD using the jet substructure became an important avenue for BSM study at LHC. 
Fig. \ref{fig:0} shows produced top quark decay (top) and quark and gluon parton shower processes (bottom two). (See Ref. \cite{Abdesselam:2010pt,Altheimer:2012mn,Altheimer:2013yza} and references therein   
). 
Recently, the jet substructure techniques evolved to 
ML-based classification algorithms for heavy particle tagging, e.g. $W,Z,h,t $ tagging
\cite{Almeida:2015jua,Butter:2017cot,Kasieczka:2017nvn, Louppe:2017ipp,Kasieczka:2019dbj, Chakraborty:2020yfc,Bhattacharya:2020aid, Ju:2020tbo,Dreyer:2020brq, Tannenwald:2020mhq,
Dreyer:2022yom,Hammad:2022lzo,Ahmed:2022hct,Munoz:2022gjq,He:2023cfc, Aguilar-Saavedra:2023pde, Athanasakos:2023fhq,Grossi:2023fqq,Hammad:2024hhm,
Hammad:2024cae}. 
and being applied to LHC analysis in ATLAS and CMS \cite{CMS:2017wtu,ATLAS:2017gpy, CMS:2020poo, ATLAS:2023gog, Andrews:2021ejw, Keicher:2023mer} They are also applied to the case with multiple heavy particle final states \cite{Baron:2023yhw,Hammad:2023sbd,Esmail:2023axd} or extraction of the fundamental parameters \cite{Datta:2019ndh,Chakraborty:2019imr,Kim:2021gtv,Subba:2022czw,Bogatskiy:2023nnw}. 
Conversely, if the process generates one or more spatially isolated quarks or gluons, the event can be reconstructed using one or more small-radius "thin" jets. This is usually followed by techniques to identify the "flavor" of the initiating quark (up, down, strange, charm, bottom) or gluon, a process known as flavor tagging. Hadrons originating from $b$ and $c$ quarks (predominantly $B$ and $D$ hadrons) have a relatively long lifetime, enabling them to travel a short distance within the detector. This results in tracks that are displaced from the primary collision point, primary vertex, PV,  and appear to originate from a displaced vertex, i.e. secondary vertex. Using ML methods, such information is combined with subsequent decay information, leading to a large enhancement of the tagging efficiency \cite{Akar:2020jti,Shlomi:2020ufi,Goto:2021wmw,Guiang:2024qzk}. The possibility to identify strange quark using ML is also discussed \cite{Erdmann:2019blf, Nakai:2020kuu, Erdmann:2020ovh}. A gluon is in octet representation in QCD SU(3) symmetry and jet initiated from gluon can also be distinguished from the other quarks in SU(3) triplet representation\cite{Komiske:2016rsd,Cheng:2017rdo,Abbas:2020khd,CMS:2022fxs,Zhang:2023czx,Goswami:2024xrx}.
Because of the page limitation, this article does not discuss the ML application of flavor tagging of narrow jets. 

The application of ML to jet tagging offers several advantages over traditional methods:

\textbf{Improved Accuracy:} 
ML algorithms can capture the complex correlations and patterns of large datasets that traditional methods find challenging. This improves tagging accuracy and better discrimination between different jet types.

\textbf{Scalability:} ML models can be trained on large datasets and then deployed to analyze new data quickly \cite{Qu:2022mxj}. This scalability is crucial in the current LHC experiments, where millions of collisions are recorded every second.

\textbf{Flexibility:} ML models can be easily adapted to new types of data or new tagging tasks. For example, a model trained on jets from one type of particle collision can be fine-tuned to work on jets from a different type of collision with minimal retraining \cite{Dreyer:2022yom,Beauchesne:2023vie}.

\textbf{Automation:} ML techniques automate the feature extraction process, reducing the reliance on manually defined jet substructure variables. This automation allows for the discovery of new features that may have been overlooked by traditional methods.

This article is organized as follows:  In Sec.\ref{sec:2}, we review the different types of data representation for jet tagging analysis. Sec.\ref{sec:3} discusses the different deep learning methods used for heavy flavor jet tagging with a primary focus on transformers. We describe several approaches to incorporate physically motivated structures in  Sec. \ref{sec:35}. 
While the generic ML technique offers significant improvement, there are still some possibilities for improving the algorithm.  The improvement comes from the network structure which ensures the physically motivated features such as Lorentz invariances and QCD structure of the jets. 
In Sec.\ref{sec:4}, we discuss different methods to interpret the network's decision-making. Our conclusion is drawn in Sec.\ref{sec:5}.
\section{ Data representation for jet tagging}  
\label{sec:2}
The initial step in any machine learning problem is the determination of the structure of the data format. Jets have a complex structure, and there isn't a single method for encoding information about their radiation patterns into a specific data structure. Early machine learning approaches to jet substructure employed a set of one-dimensional, physically-motivated observables.  However, this method doesn't ensure that all available information is accessible to the neural network, raising the need for a better representation of the jet information. 

\subsection{Image based dataset}
\begin{figure}
    \centering
    \includegraphics[width=\linewidth]{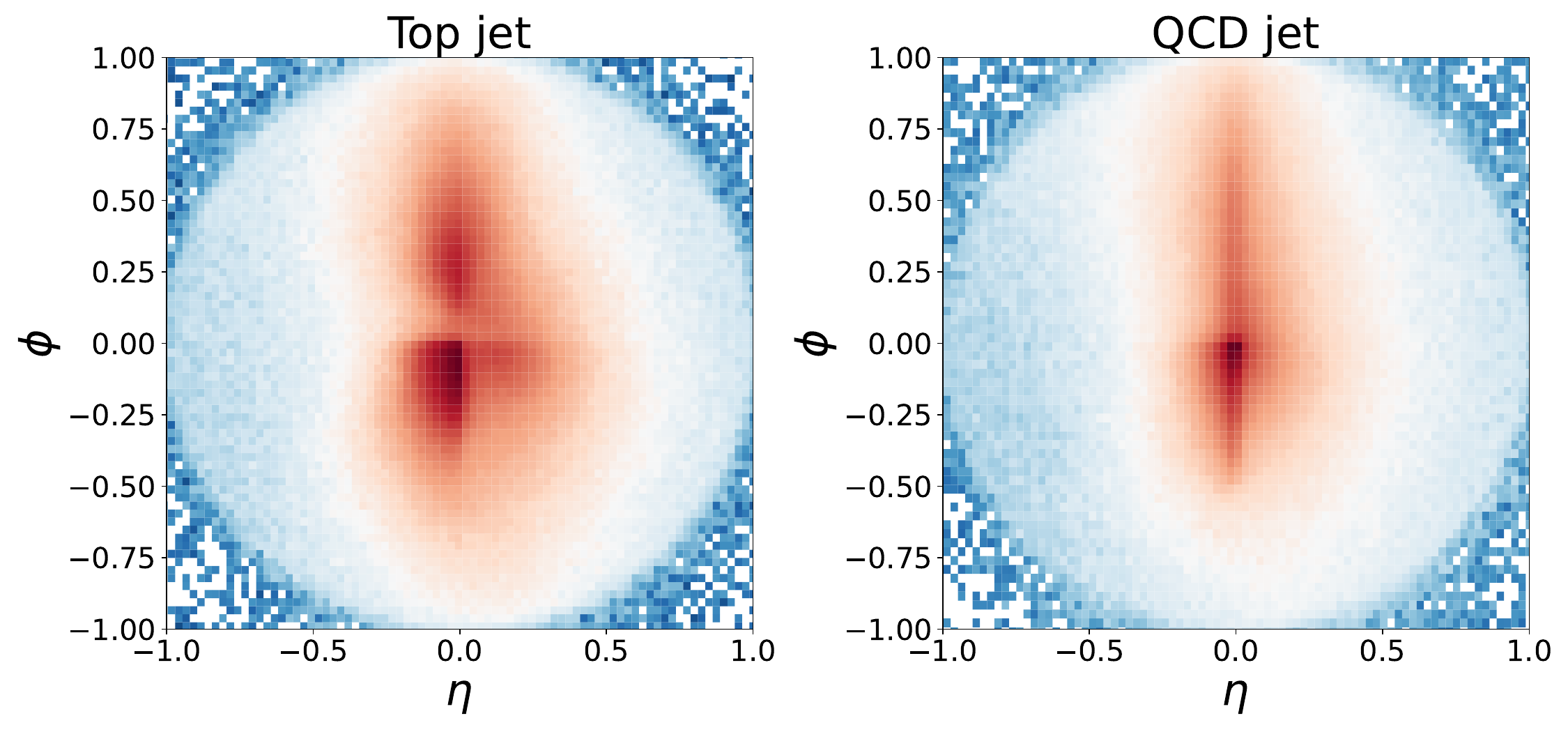}
    \caption{Image for top jet (left) and QCD jet (right) after the preprocessing steps of centering, rotation and flipping.}
    \label{fig:1}
\end{figure}
In the recent years applications of deep neural networks (DNNs)  analyze the  
jets information as images \cite{Cogan:2014oua,deOliveira:2015xxd,Barnard:2016qma,Komiske:2018oaa,Lee:2019cad,Collado:2020ehf,Li:2020bvf,Li:2020grn,Filipek:2021qbe,Han:2023djl,Kheddar:2024osf}. 
A jet image is a pixelated grayscale image, where the pixel intensity corresponds to the energy (or transverse momentum) of particles depositing energy in a specific detector location. Figure \ref{fig:1} shows an example of a top jet image (left) and a QCD jet image (right). The color bar represents a cumulative $P_T$ distribution of $50000$ events of both jets.  The convolutional neural network (CNN) is the natural machine learning tool for analyzing images. Jet images have several characteristics that make them suitable for jet flavor tagging techniques. First, each image maintains a consistent dimensionality, ensuring that every jet can be described by a fixed set of numbers. 
Second, the CNN filters 
easily capture characteristic features of the jet substructure. 
However, they still face challenges such as image sparsity and difficulties in incorporating additional features, like particle ID.

\subsection{Graph based dataset}
To address the issues in image-based DNNs, one effective method is to use a graphical structure consisting of nodes and edges to encode particle information.  A graph is minimally defined by a set of nodes and a set of edges connecting pairs of nodes. Graph Neural Networks (GNNs) can then be employed to incorporate the topological relationships among the nodes and edges to  effectively learn from graph-structured data \cite{Abdughani:2018wrw, Moreno:2019bmu,Bernreuther:2020vhm,Iiyama:2020wap,Heintz:2020soy,
ExaTrkX:2020nyf, Dreyer:2020brq,
Pata:2021oez,Verma:2021ceh,Atkinson:2021jnj,Gong:2022lye,Ma:2022bvt,Bogatskiy:2022czk,Builtjes:2022usj,DiBello:2022iwf,Mokhtar:2022pwm,Huang:2023ssr,Duperrin:2023elp,Konar:2023ptv}. 

GNNs are a class of deep learning architectures designed to operate on graph-structured data, implementing strong relational inductive biases. They use parameterized message-passing to propagate information across the graph, allowing for sophisticated edge, node, and graph level outputs. Within a GNN, standard neural network building blocks, typically fully connected layers, are used to implement the message computations and propagation functions. %

In many jet tagging tasks, the relationships between different hadrons are not straightforward. Therefore, a decision must be made on how to construct a graph from the set of inputs. 
Edges in the graph serve as communication channels among the 
nodes and indicate the relationship between objects. The variables associated with nodes and edges are updated in the message passing, allowing to construct variables most suitable for the given tasks.

As the number of edges between other nodes increases, the computational load to create an edge representation becomes prohibitive. 
Therefore,  
it is necessary to 
limit the number of edges contact with a node. 
Edges can be formed based on a relevant metric, such as the angles between the particles $\Delta R$ reconstructed by a detector or the physical distance between detector modules. 
In addition, one can directly limit the number of nodes, such as 
connecting only k-nearest neighbors.

Node features used to connect edges can also be based on what is learned by the networks.  For example, the distance between nodes 
may be updated in message passing, and a new diagram can be constructed using the distance measure. This approach sometimes referred to as dynamic graph construction, is used in architectures like EdgeConv and GravNet\cite{Qasim:2019otl}. Even in such a case, the gradient of the neural network parameters is affected only by those nodes that are initially connected.

\subsection{Particle cloud dataset}
For jet physics, 
an unordered, permutation-invariant set of particles is a more intuitive representation than jet images or graphs. This form of representation, known as a particle cloud, retains all the benefits of particle-based models, particularly the flexibility to incorporate arbitrary features for each particle. Like the point cloud representation used in computer vision, particle clouds provide a robust framework for modeling jets.

Graph or image representations 
require particles to be sorted, thereby the core part of the network relies on the order of particles. Particle clouds do not impose any order. Since the particles within a jet lack intrinsic order, any manually imposed arrangement may be suboptimal and reduce performance. 

Several design principles and architectural strategies can be employed to ensure permutation symmetry in networks. The goal is to construct the network so that the output remains unchanged regardless of the order of the input elements.

The most straightforward approach is to use symmetric functions within the network. A function $( f )$ is symmetric if $ f(x_1, x_2, \ldots, x_n) = f(x_{\sigma(1)}, x_{\sigma(2)}, \ldots, x_{\sigma(n)}) $ for any permutation $( \sigma )$ of the indices. Examples of symmetric functions include summation, mean, and maximum \cite{qi2017pointnet}. 

The Deep Sets framework \cite{zaheer2017deep,Komiske:2018cqr,zhang2022set}, provides a principle method to build permutation invariant networks. According to this framework, any permutation invariant function can be decomposed into two components: a transformation applied to each element of the set independently and a symmetric aggregation function applied to the transformed elements. Mathematically, this can be expressed as:
\begin{equation}
    f(X) = \rho \left( \sum_{x \in X} \phi(x) \right)
\end{equation}
where $( \phi )$ is a function applied to each element $( x )$ in the set $( X )$, and $( \rho )$ is a symmetric function, typically a summation. This decomposition ensures that the network is permutation invariant.
GNNs naturally extend to permutation invariance when the graph representation of the set and a symmetric function is used. 
The message-passing mechanism in GNNs aggregates information from neighboring nodes in a permutation invariant manner. By using the GNNs for sets, one can leverage the expressive power of graphs while ensuring permutation invariance.

Transformer networks are a class of models that are intrinsically permutation invariant. The core of 
the transformer model is the attention matrix. The attention matrix is a weighted sum of input elements, where the weights are determined by the relevance of each element to the others. By ensuring that the attention weights are computed in an invariant permutation way, the resulting network can process sets effectively. 

Mathematically, the attention mechanism works as follows: Given the input to the multi-head attention as $X_{i,j}$  representing one of the input datasets while $S_{n,m}$ is the other data set which you want to calculate the correlation.  If $S$ is equal to $X$ it is called self-attention. We then build, 
\begin{equation}
\begin{split}
    &Q^{i \times j} \equiv X^{i \times j} \cdot W_Q^{j \times j}\,, \hspace{4mm} K^{n \times j} \equiv S^{n \times m} \cdot W_K^{m \times j} \,, \\
    &\hspace{20mm} V^{n \times j} \equiv S^{n \times m} \cdot W_V^{m \times j} \,,
\end{split}
\end{equation}
where $ Q, K $, and $ V $ are the query, key, and value matrices, respectively, and will be used to compute the attention of the dataset. The superscripts indicate the dimensions of the matrix. The scaled dot product attention score $\alpha$ is defined as:

\begin{equation}
\begin{split}
    \alpha^{i \times n} &\equiv\text{softmax}\left(\frac{Q^{i \times j} \cdot (K^{n \times j})^T}{\sqrt{d}} \right) \\
    &= \frac{\text{exp}(Q^{i \times j} \cdot (K^{n \times j})^T / \sqrt{d})}{\sum_i \text{exp}(Q^{i \times j} \cdot (K^{n \times j})^T / \sqrt{d})}\,.
\end{split}
\end{equation}

The resulting attention weight matrix has the dimensions of the first input dataset, $X_{i,j}$. The attention output is obtained as:
\begin{equation}
    \mathcal{Z}^{i \times j} = \alpha^{i \times n} \cdot V^{n \times j}\,.
\end{equation}
The attention output matrix has the same dimensions as the first input dataset $ X_{i,j}$.
each transformer layer contains a multi-head attention mechanism, which integrates different attention heads to enable parallel, multi-dimensional processing of inputs. 
This allows the model to simultaneously perceive various aspects of the input sequence, facilitating the understanding of diverse and subtle connections within the data and providing a more comprehensive representation.

These attentions are then linearly combined to produce the final output of the multi-head attention layer. The output of the multi-head cross-attention is given by:
\begin{equation}
    \mathcal{O}^{i \times j} = \text{concat}\left(\mathcal{Z}^{i \times j}_1, \mathcal{Z}^{i \times j}_2, \ldots, \mathcal{Z}^{i \times j}_n \right) W^{(n \cdot j \times j)}\,,
\end{equation}
where \( W^{(n \cdot j \times j)} \) is the learnable linear transformation matrix that retains the dimensions of the input dataset. 
The attention output is used to scale the input dataset via a skip connection as:
\begin{equation}
    \widetilde{X}^{i \times j} = X^{i \times j} + \mathcal{O}^{i \times j}\,.
\end{equation}
The transformed dataset $\widetilde{X} $ signifies the importance of each element relative to all elements within the set.

\section{Application of Deep-Learning to particle physics}
\label{sec:3}

Several machine learning models using particle clouds have been introduced for collider analysis, such as  Deep Sets \cite{Komiske:2018cqr}, Edge Convolution (EDGCNN) \cite{Qu:2019gqs}, and Transformers \cite{
Shmakov:2021qdz,
Qu:2022mxj,Finke:2023veq,
Hammad:2023sbd,He:2023cfc}. 
The Deep Sets model,
achieves state-of-the-art performance but requires a large latent space, resulting in increased complexity. The Edge Convolution Neural Network (EDGCNN) incorporates local information from the nearest neighbors of each particle to enhance learning. Other models, such as JEDI-net \cite{Moreno:2019bmu}, Point Cloud Transformer \cite{Mikuni:2021pou}, Lorentz Net \cite{Gong:2022lye}, and PELICAN \cite{Bogatskiy:2022czk}, are also employed in particle cloud analysis.

The particle Transformer \cite{Qu:2022mxj} is a transformer-based model specifically designed for particle cloud analysis. 
This mechanism calculates self-attention weights among all pairs of particles in the set.  Typically, up to n-th particles ordered by $p_T$ are included in the set. The operation within the network maintains the permutation invariance of the input token order. 

Transformer models have demonstrated superior performance in collider analysis but are characterized by high model complexity and significant runtime requirements.

Transformers were originally introduced as sequence-to-sequence models for machine translation, with their core architecture centered around the encoder-decoder block.
In the context of the LHC event classification analysis, models that utilize only the encoder block.
The transformer encoders have been adapted to process particle clouds, representing final state particles as a permutation-invariant sequence.
The primary motivation for using transformer encoders with particle clouds is their inherent ability to model interactions between particles regardless of their spatial proximity. 

Particle transformer network \cite{Qu:2022mxj} extends the basic self-attention mechanism by imposing pairwise information as 
\begin{equation}
    \text{P-MHA}(Q,K,V)  = {\rm softmax} \left(Q\cdot K^T/\sqrt{d} + U \right)\cdot V\,,
\end{equation}
with $U$ is the pairwise feature between any two hadrons in the dataset. This enables P-MHA to integrate particle interaction features derived from physics principles and adjust the dot-product attention weights, thereby enhancing the expressiveness of the attention mechanism.

\section{physics application}
\label{sec:35}
\subsection{Integrating jet physics with the deep learning }

As briefly mentioned in the introduction, the physics process obeys several essential features, such as symmetry and scales. Deep learning can, in principle, find such structures from the huge dataset, but incorporating the physics structure in the networks is rewarding. Permutation invariance discussed in the previous sections is an example of such a structure.  In this subsection, we describe some of the recent attempts in this direction.

 PELICAN(Permutation Equivariant and Lorentz Invariant or Covariant Aggregator Network for Particle Physics) \cite{Bogatskiy:2022czk} is a permutation-equivariant network that remains invariant under Lorentz transformations. Its main equivariant block consists of a simple dense layer and an aggregation block. The aggregation block employs 15 linear aggregation functions, each multiplied by a trainable exponent \(\alpha\) whose length matches the number of hadrons in the dataset. This setup allows for flexibility in the aggregation process.{For instance, setting \(\alpha = 1\) results in a sum aggregation function. } Combining multiple aggregators has been shown to enhance jet tagging accuracy. 

Importantly, the PELICAN structure has a small number of parameters compared with earlier Lorentz invariant networks such as LorentzNet\cite{Gong:2022lye} ensuring low computational cost. Lorentz invariant network has been actively studied recently \cite{Spinner:2024hjm, Bhardwaj:2024djv}

In the hadron collider, the particles going the same direction are clustered into a jet based on the distance between particles or pseudojets(clusters of particles).  
LundNet \cite{Dreyer:2020brq} utilizes the clustering sequence of jet reconstruction to construct a graph. 
The clustering sequence is infrared and collinear (IRC) and safe in QCD radiative correction. LundNet utilizes the clustering process.  The distance between the pseudojet is defined as follows. 
\begin{equation}
d_{ij}=z_i^k z_j^k \Delta R
\end{equation}
Here the $R$is the angle between two pseudjet defined as $\Delta R= \sqrt{(\Delta \phi)^2 + (\Delta\eta)^2}$ where $\phi$ is azimuth, and psudorapidity $\eta=-\ln (\tan (\theta/2))$ where $\theta$ is inclination. 
The $z_i$ is a fraction of a pseudojet energy $E_i$,  $z_i = E_i/E_j$. Depending on the purpose of the physics, the appropriate value ($k=-1,0, 1$) is selected. 
The jet clustering algorithm marge the pair of the pseudojet $i$ and $j$ with the smallest distance,  creating a new pseudojet with momentum $p=p_i+p_j$ until ${\rm min}(\Delta R)> R_{cut}$.  The jets are the final pseudo jets, and particles consisting of a jet called the jet constituents. 
The transverse momentum scale of $p_{i(j)}$ relative to $p$, $k_T$ specifies the physical scale of the pseudojet marging. 

In LundNet, one uses the backward clustering history to create a graph. The graph involves nodes with three neighbours only, where two edges correspond to the two pseudojets and one edge corresponds to the marged pseudojet.  
The node information is the kinematical information of the splitting. The kinematical information can be based on the 2 dim plane called the Lund plane, where the name of the network comes from. 
The proposed network shows higher performance compared with ParticleNet, and the required computing resources are significantly smaller. 
The LundNet classification performs better if the cut on the clustering scale $k_T$ is smaller. This shows the importance of the correct theoretical prediction of parton showers up to a smaller splitting scale.  Recent next-next-to-leading-log (NNLL) improvement of parton showers aims toward the accurate parton shower to achieve the target(See Ref. \cite{vanBeekveld:2024wws} and references therein).

\begin{figure}[t]
    \includegraphics[width=\linewidth]{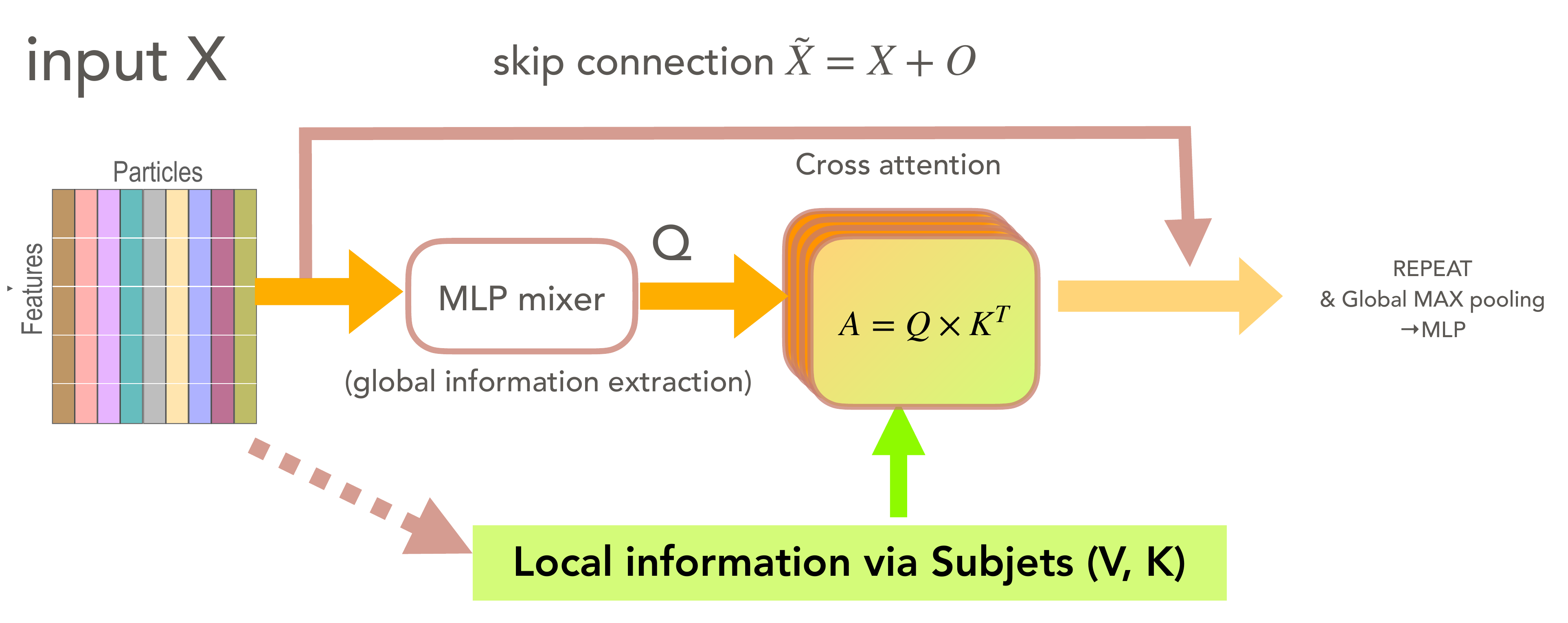}
    \caption{A Schematical figure of the Cross Atention-Mixer network. Input X is the $n$ jet constituents (particles) with the feature of size $d$. 
    They are processed by two MLPs which mix feature and particles subsequently. Input X is also used to reconstruct subjets. Cross attention is calculated between jet constituents and subjets. }
    \label{fig:CAMixer}
\end{figure}

In Ref. \cite{Hammad:2024cae}, we have proposed another utilization of the jet clustering. 
For the given jet, we can take the jet constituents and recluster them using the jet clustering algorithm using smaller $R_{cut}$. 
The subjets carry information on the wide-angle splitting of the parton shower, where each parton carries significant momentum and is reasonably localized. 
We use them as the second information of the transformer, $S_{ij}$, while  $X_{ij}$ arise from the whole constituents of the jet. 
The $Q$ and  $K$ are then attention between different objects and called cross attention. 
The network to process $X$ before going into cross attention can be replaced from the transformer to significantly small networks called  MLP-mixer, which focus on the jet-wide global features; the localized cluster information is encoded by subjets. The mixer network consists of the two MLPs, which mix the particle and feature tokens, respectively. In the following, we call the whole network a Cross-Attention-Mixer network(CA-Mixer network). See Figure \ref{fig:CAMixer} for a schematical network description.  

The CA-Mixer network keeps the size of the initial input as the self-attention block but is significantly smaller both in size and in computational cost \cite{Hammad:2024cae}. 
The mixer network is a permutation-invariant network, and the subjet inputs are also permutationally invariant because jet clustering algorithms are designed to be permutation-invariant. The cross-attention outputs go through the summation and global pooling carefully to make the final output also permutation-invariant.

The cross-attention network is suited to study the correlation between hard partons and hadrons. 
In the figure, quarks from the hard process produce relatively hard partons through the parton shower. The hard parton momenta are correlated with the subjet momenta due to the factorization of QCD. In this context, we want to analyse the hadron distribution conditioned by resolved partons arising from parton shower process, while the parton momentum is expected to be close to the subjet momentum. 
Due to the correlation between parton momenta and (sub)jet momenta, the QCD process may schematically be expressed as 
\begin{equation}
\prod_k P_s\left(  \ \{x_k\}_{J_k}\ \vert  \{J_i\} \  \right)P_h\left(  \ \{J_i\}\ \right)  \,,
\label{eq:facii}
\end{equation}
where $P_s\left(  \ \{x_k\}_{J_k}\ \vert  \{J_i\} \  \right)$ is the hadron distributions in $J_k$, conditioned by the jets in the process, and $P_h$ is the distribution of jets $\left\{J_i\right\}$, correlated with the hard process or parton shower. Inside the jet, there are similar correlations of subjets replacing jet  $\rightarrow$ subjets. 
Note that $P_s$ is in principle conditioned by all hard activities in the events(or subjets in the jet)  due to the long range correlations called color coherence. In the network, this long range correlation is taken into account in the Mixer network. 

The cross attention can be extended to study not only for a jet substructure but also for the analysis of the event which contains several fat jets.  
In Ref. \cite{Hammad:2023sbd}, the cross attention mechanism is 
utilized for the two Higgs boson final state arising from a heavy Higgs boson ($pp\rightarrow H\rightarrow hh$), where a high background rejection is achieved by utilizing the cross attention of jet constituent of two fat jets and the jet momenta. 

\subsection{Top tagging}  
\begin{table*}[th!]
\centering
\caption{ Taken from Ref. \cite{Hammad:2024cae}.  Performance of the CA-Mixer network for top quark tagging compared with other models. Results for JEDI-net \cite{Moreno:2019bmu}, Point Cloud Transformer (PCT) \cite{Mikuni:2021pou}, Lorentz Net \cite{Gong:2022lye}, PELICAN \cite{Bogatskiy:2022czk}, PFN \cite{Komiske:2018cqr}, ParticleNET \cite{Qu:2019gqs}, and ParT \cite{Qu:2022mxj} are quoted from their published results. Pretrained ParticleNET and ParT have higher performance with AUC = $0.9866$ and AUC= $0.9877$, respectively. The pertaining is done on the JETCLASS dataset, followed by the tuning to the top dataset.  Training time is per epoch with a batch size of $1024$.  The GPU training time is measured on an NVIDIA RTX A6000 card. 
}
\label{tab:1}
\begin{tabular}{cccccc}
\hline
&AUC & Rej$_{50\%}$ & Parameters & Time (GPU) [s] \\
\hline
JEDI-net with $\sum \mathcal{O}$ & $0.9807 $  &$-$& $87.7$K& $-$         \\
\hline
 PFN & $0.9819$  &$247\pm$ 3& $86.1$K& $\mathbf{30}$ \\
\hline
PCT & $0.9855$  &$392\pm$ 7& $193.3$K& $-$   \\
\hline
LorentzNet & $0.9868$  &$498 \pm $18& $224$K& $-$           \\
\hline
ParticleNET   & $0.9858$  &$397\pm 7$& $370$K& $-$             \\
    \hline
     PELICAN            & $\mathbf{0.9869}$  &$-$& $\mathbf{45}$K& $-$             \\
    \hline
    ParT                   & $0.9858$  & $413\pm 16$   &$2.14$M& $612$          \\
    \hline
CA-Mixer         & $0.9859$  & $416\pm 5$  & $86.03$K& $33$          \\
    \hline
\end{tabular}
\end{table*}

In this subsection, we compare the performance of the networks described in section 4.1 for top vs QCD classification. 

The top quark strongly couples to Higgs bosons, and the top Yukawa coupling affects vacuum stability. Searching for deviation from the standard model in the top and Higgs sectors is currently one of the most important topics in high-energy collider physics. 
The top quark decays into $b$ quark and $W$ boson, and $W$ boson decays into either quark-anti quark or charged lepton and neutrino.  The boosted top quark decays fully hadronically ( bottom quark + quark + anti-quark) and can be identified with high efficiency using Deep learning.

To evaluate the performance of the proposed networks, we use the simulated top tagging dataset as described in Ref. \cite{Butter:2017cot}. Jets in this dataset are produced at a center-of-mass energy of $( \sqrt{s}=14 )$ TeV using Pythia8 \cite{Bierlich:2022pfr}. 
Fast detector simulation is performed with Delphes \cite{deFavereau:2013fsa}, excluding multiple parton interactions and pileup effects. 
Jets are clustered from Delphes E-Flow objects using the Anti-kt algorithm with a radius parameter $( R=0.8 )$. 
It should be kept in mind that the data is a correction of detector hits in the actual experimental application, which went through the subsequent object reconstruction. Systematic errors in the measurements might significantly affect the final results. 

Jets are selected based on their transverse momentum $( p_T \in [550,650] )$ GeV and pseudorapidity $( |\eta| < 2 )$. For top quark events, a jet must be within $( \Delta R = 0.8 )$ of a hadronically decaying top quark, with all three decay products within $( \Delta R=0.8 )$ of the jet axis. The background consists of QCD dijet processes.
The dataset comprises a million of  $t\bar{t} $ events and a million QCD dijet events. 
This dataset is widely used in the literature, facilitating performance comparisons with other networks.

Table \ref{tab:1} presents networks performance metrics, including the number of tunable parameters and training time per epoch. Bold numbers highlight the best performance of all the considered networks. 

Among the network model, the Cross-Attention-Mixer Network(CA-Mixer) is able to achieve a comparable performance as the Particle transformer model(ParT).
Achieving this level of performance requires both jet constituent information processed by the Mixer network and subjet inputs included through cross atention.  The CA-Mixer not only achieves state-of-the-art performance comparable to ParticleNET and ParT but also trains approximately 20 times faster. PFN, while having the shortest training time, cannot effectively learn local information shared between particles and their neighbors, resulting in relatively poorer performance.  PELICAN network has the best tagging performance with the lowest computational cost, because the network is equivaliant and the parameter space that network survey is small. CA-Mixer networks achieve improvement in computational performance based on conceptually different ideas.  Further improvement might be possible by 
introducing Lorentz equivalence. CA-Mixer network has a simple data structure stacking all information; therefore, extending it to have Lorentz equivariance requires careful consideration of the data structure. 

\section{Interpretation AI methods}
\label{sec:4}
Interpretable AI methods to elucidate network decision-making enhance both the reliability and usability of AI systems. Transparency of the decision-making process is paramount, fostering trust among users and stakeholders.
Interpretability also aids in debugging and error correction, enabling physicists to identify and rectify flaws or biases within the model, thereby enhancing its accuracy and robustness.
Moreover, interpretable methods help to identify the critical physical quantity contributing to the network decision. 
Overall, incorporating interpretable AI methods is essential for developing reliable 
and effective AI applications that drive the next innovation and enhance decision-making across diverse domains. 
The ultimate hope is to find novel concepts from deep learning applications which improve our insights into physics. 

In this section, we explore 
methods for interpreting and analyzing the classification results. 
Commonly employed interpretation methods include:

\textbf{Central Kernel Alignment (CKA)}: CKA serves as a metric to assess the similarity between two sets of learned representations within a high-dimensional feature space. Initially introduced in Ref. \cite{kornblith2019similarity} and applied in collider analysis in Ref. \cite{Esmail:2023axd}, CKA evaluates how network layers or hidden layers from different models capture both local similarities and global structure.

\textbf{Grad-CAM}: This technique generates class-specific activation maps by weighting the gradients of the predicted class score with respect to the final transformer layer. Grad-CAM highlights regions in the feature space that are crucial for the model’s classification decision, offering a geometrical interpretation in the $\eta-\phi$ plane of the information learned by the network.

\textbf{Saliency Maps}: Saliency maps visualize the importance of different parts of the input sequence concerning the model’s predictions. They highlight the regions of the input that most significantly influence the model’s output, providing insights into the model’s decision-making process.

\textbf{Attention Maps}: Particle transformer 
use the attention matrix. The attention map highlights the correlation of the particles in the cloud that receive higher attention and are deemed most relevant for prediction, facilitating an intuitive understanding of the model's decision-making process.

The following describes the methods discussed in more detail: CKA, attention maps, and Grad-CAM. We will apply some of the methods for the transformer encoder with cross-attention introduced in section 4, aiming to illuminate their outcomes. These interpretation methods are generalizable and applicable to other networks. 

\subsection{CKA similarity}  

One of the primary challenges in analyzing hidden layer representations in neural networks is the dispersion of features across neurons, which can vary significantly across layers and models. 
CKA addresses this challenge by enabling quantitative comparisons within individual networks and across different models. 
CKA rooted in kernel methods and alignment-based metrics, provides a robust framework for assessing similarity between learned representations from different models or layers within a model. 
It evaluates the alignment between activation matrices of two hidden layers, $X $ and $ Y $ evaluated on the same input dataset, where $ X \text{ in } \mathbb{R}^{d\times P_1} $ and $ Y \text{ in } \mathbb{R}^{d\times P_2}  $, with $ d $ denoting the data size, and $ P_1 $ and $ P_2 $ representing the number of neurons in each layer. The CKA similarity is defined by:
\begin{equation}
    \text{CKA}(M,N) = \frac{\text{HSIC}(M,N)}{\sqrt{\text{HSIC}(M,M) \text{HSIC}(N,N)}},
\end{equation}
where $ M = XX^T $ and $ N = YY^T $ are Gram matrices of the two hidden layers with dimensions $ d \times d $ and has same size for all hidden layers. The Hilbert-Schmidt Independent Criterion (HSIC) between matrices $M $ and $N $ is expressed as:
\begin{equation}
    \text{HSIC(M,N)} = \frac{1}{(d-1)^2} \text{Tr}(MHNH)\,,
\end{equation}
with $ H $ being a centering matrix, $ H_{ij} = \delta_{ij} - 1/d $. Centering the matrices ensures that CKA is robust to outliers and extreme values in the data, thus enhancing the reliability of comparisons between representations.

\begin{figure}[h!]
    \includegraphics[width=\linewidth]{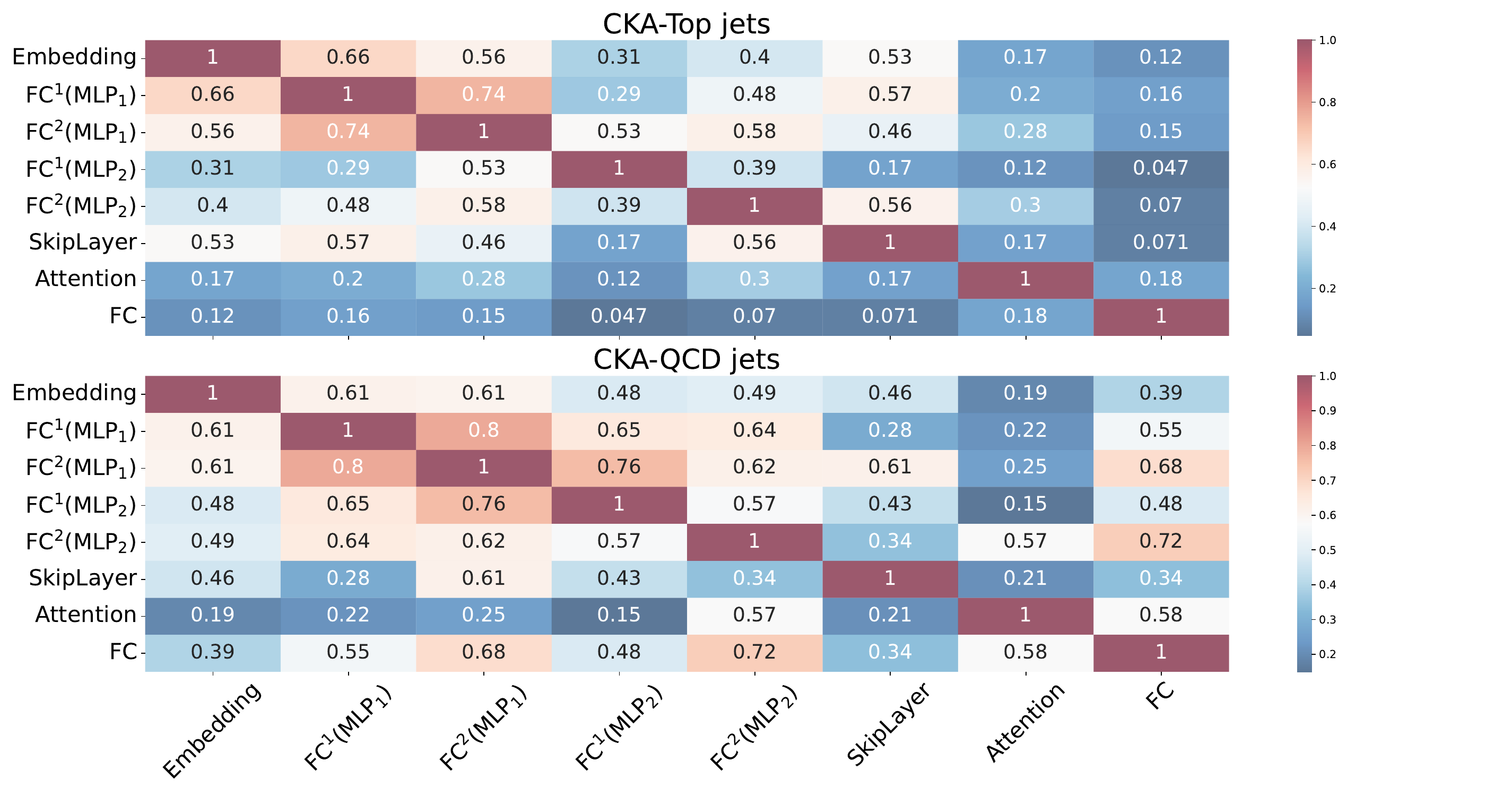}
    \caption{ Taken from Ref. \cite{Hammad:2024cae}. CKA similarity of top jet events (top plot)
and QCD jet (bottom). 
Axes represent the network layers.
FC$^{1(2)}$(MLP1) and FC$^{1(2)}$(MLP2) are the two fully connected layers that mix the particle and features, respectively. 
Attention is the multi-heads cross-attention. FC represents the last FC layer in the whole network, }
    \label{fig:3}
\end{figure}
CKA values range from $0$ to $1$, where higher values indicate layers capturing redundant information, and lower values suggest distinct or complementary information. This insight into layer similarity aids in understanding model performance and suggests opportunities for model optimization or simplification based on representation redundancy.

To highlight the importance of the CKA similarity, we adopt the CKA to interpret the similarity of the hidden layers representation of the Mixer network when it is used to distinguish the top from QCD jets \cite{Hammad:2024cae}. Figure \ref{fig:3} shows the CKA values for all the network layers with top jet events in the upper plot and QCD jet events in the lower plot. The multi-head cross-attention layer shows lower similarity with the two MLPs for the top jet and for the QCD jets, which indicates the importance of incorporating the cross-attention layer into the network.  Lower CKA  values for the top jet, specifically in the FC and attention layers,  indicate that the network layers are adept at capturing distinct information and are capable of learning the substructure of the top jet.
\subsection{Attention maps}  
The attention weights are a core of transformer models. 
Attention maps visualize the attention scores allocated to individual particle tokens within the input sequence, offering insights into where the model directs its focus during processing. 
Analyzing attention maps identifies particle tokens receiving higher attention scores, signifying their importance in the model's decision-making process. The attention matrix is calculated from pairs of particle tokens; therefore, the attention maps are also sensitive to how they interact with each other. 
It acts as a link between the abstract computations performed by neural networks and the need for interpretability.

Each element of  $\alpha_{ij}$ in Eq. 3  represents the strength of the attention from particle $x_i$ to particle $x_j$.
The attention weights $ \alpha_{ij}$ are calculated for a test data set of size $d$ to visualise the attention maps. It can be reshaped into a matrix form by taking the average of the same elements.  This matrix can be plotted or visualized in 2D or 3D space, depending on the dimensionality of the particle cloud. High values of $\alpha_{ij}$ indicate that particle token $x_i$ heavily relies on particle tokens $x_j$ for information, suggesting a strong local relationship. On the other hand, Uniform or distributed $\alpha_{ij}$ values across all index $j$ indicate that particle token $x_i$ considers all points in the dataset equally, suggesting a global understanding or context.

\subsection{ Grad-Cam}  
Grad-CAM \cite{selvaraju2020grad} is an approach aimed at visualizing and interpreting the decisions made by deep neural network (DNN) models. It expands upon the concept of class activation maps (CAMs) \cite{zhou2016learning}, adapting it to models with diverse architectures. In the context of the transformer model for jet classification, 
Grad-CAM 
highlight the regions contributing to the 
prediction of a specific class 
in the transformed feature space, specifically the $\widetilde{\eta}-\widetilde{\phi}$ plane. In the transformer model, the location of the jet constituent $\eta$ and $\phi$ is a particle feature and the subject of the transformation of Eq. 6. It still maintains the relation with the original coordinate by skip connection.

Let  $F_{k}=\widetilde{\eta}$ and $\widetilde{\phi}$,
denote the activation of the final transformer layer(namely, transformed coordinates) of  a particle  of the $k^{\text{th}}$ event. 
The gradient of the predicted class score ($Y_c$) with respect to $i$-th constituent of the jet  can be computed as 
\begin{equation}
G_{k,i}=\frac{\partial Y_c}{\partial F_{k,i}}\,,%
\end{equation}
where $i$ indicates the suffix of different outputs. 
These gradients are globally summed over all jet constituents to derive importance weights ($\alpha$):
\begin{equation}
\alpha_k
= \frac{1}{Z} \sum_{i=1}^{2N_c } G_{k,i}.
\end{equation}
Here, $( Z)$ represents the size of the feature activations, and the sum is taken over all the jet constituents of the size $N_c$.  
The final Grad-CAM heatmap for $n$ events is computed as:
\begin{equation}
\text{Grad-CAM}(\widetilde{\eta},
\widetilde{\phi} )
= \frac{1}{n} \sum_{k=1}^{n}
\alpha_k
\widetilde{p_T}(\widetilde{\eta}, \widetilde{\phi}), 
\end{equation}
where $\widetilde{p_T}(\widetilde{\eta}, \widetilde{\phi}) $ is the sum of the transformed $p_T$ at
$(\widetilde{\eta},\widetilde{\phi}) $

This heatmap visually emphasizes the regions within the input image that contribute most significantly to predicting the target class.

While Grad-CAM effectively explains feature space regions relevant to network predictions, its reliance on gradient information from the final transformer layer may limit its ability to capture long-range dependencies crucial for decision-making. Additionally, Grad-CAM's sensitivity to input perturbations poses challenges in scenarios involving adversarial examples, potentially compromising its robustness.
\section{Conclusion}  
\label{sec:5}
In this article, we review recent deep learning methods for heavy flavor jet tagging, with a primary focus on attention-based transformer networks. We discuss various data structures used by different networks, highlighting the importance of the particle cloud representation over structures such as jet images or graphs. The main advantage of the particle cloud representation is its permutation invariance property. Transformer networks are naturally permutation invariant and are well-suited for particle cloud analysis. On the other hand, other networks, such as GNNs or CNNs, can achieve permutation invariance if a symmetry function is imposed.

We also discuss the importance of reflecting the physically motivated structure in the network. We discuss the two cases, Lorentz invariance and QCD factorization, which both reduce the required computational resources drastically while maintaining very high performance. 

Finally, we explore different methods for interpreting network decision-making in jet tagging analysis. The intricate structure of jets requires sophisticated deep neural networks to learn the inner structure of different jet flavor. Evaluating network performance solely based on classification output may lead to incorrect predictions when tested on real data. Accordingly, understanding the numerical flow of deep neural networks is crucial to ensure that the network can learn the physical properties hidden in the given data.
\section*{Acknowledgments}
 This work is funded by grant number 22H05113, ``Foundation of Machine Learning Physics'', Grant in Aid for Transformative Research Areas and 22K03626, Grant-in-Aid for Scientific Research (C).
\bibliographystyle{JHEP}
\bibliography{biblo}

\providecommand{\href}[2]{#2}\begingroup\raggedright\begin{thebibliography}{10}

\bibitem{Butterworth:2008iy}
J.~M. Butterworth, A.~R. Davison, M.~Rubin and G.~P. Salam, \emph{{Jet
  substructure as a new Higgs search channel at the LHC}},
  \href{https://doi.org/10.1103/PhysRevLett.100.242001}{\emph{Phys. Rev. Lett.}
  {\bfseries 100} (2008) 242001}
  [\href{https://arxiv.org/abs/0802.2470}{{\ttfamily 0802.2470}}].

\bibitem{Abdesselam:2010pt}
A.~Abdesselam et~al., \emph{{Boosted Objects: A Probe of Beyond the Standard
  Model Physics}},
  \href{https://doi.org/10.1140/epjc/s10052-011-1661-y}{\emph{Eur. Phys. J. C}
  {\bfseries 71} (2011) 1661}
  [\href{https://arxiv.org/abs/1012.5412}{{\ttfamily 1012.5412}}].

\bibitem{Altheimer:2012mn}
A.~Altheimer et~al., \emph{{Jet Substructure at the Tevatron and LHC: New
  Results, New Tools, New Benchmarks}},
  \href{https://doi.org/10.1088/0954-3899/39/6/063001}{\emph{J. Phys. G}
  {\bfseries 39} (2012) 063001}
  [\href{https://arxiv.org/abs/1201.0008}{{\ttfamily 1201.0008}}].

\bibitem{Altheimer:2013yza}
A.~Altheimer et~al., \emph{{Boosted Objects and Jet Substructure at the LHC.
  Report of BOOST2012, held at IFIC Valencia, 23rd-27th of July 2012}},
  \href{https://doi.org/10.1140/epjc/s10052-014-2792-8}{\emph{Eur. Phys. J. C}
  {\bfseries 74} (2014) 2792}
  [\href{https://arxiv.org/abs/1311.2708}{{\ttfamily 1311.2708}}].

\bibitem{Almeida:2015jua}
L.~G. Almeida, M.~Backovi\'c, M.~Cliche, S.~J. Lee and M.~Perelstein,
  \emph{{Playing Tag with ANN: Boosted Top Identification with Pattern
  Recognition}}, \href{https://doi.org/10.1007/JHEP07(2015)086}{\emph{JHEP}
  {\bfseries 07} (2015) 086}
  [\href{https://arxiv.org/abs/1501.05968}{{\ttfamily 1501.05968}}].

\bibitem{Butter:2017cot}
A.~Butter, G.~Kasieczka, T.~Plehn and M.~Russell, \emph{{Deep-learned Top
  Tagging with a Lorentz Layer}},
  \href{https://doi.org/10.21468/SciPostPhys.5.3.028}{\emph{SciPost Phys.}
  {\bfseries 5} (2018) 028} [\href{https://arxiv.org/abs/1707.08966}{{\ttfamily
  1707.08966}}].

\bibitem{Kasieczka:2017nvn}
G.~Kasieczka, T.~Plehn, M.~Russell and T.~Schell, \emph{{Deep-learning Top
  Taggers or The End of QCD?}},
  \href{https://doi.org/10.1007/JHEP05(2017)006}{\emph{JHEP} {\bfseries 05}
  (2017) 006} [\href{https://arxiv.org/abs/1701.08784}{{\ttfamily
  1701.08784}}].

\bibitem{Louppe:2017ipp}
G.~Louppe, K.~Cho, C.~Becot and K.~Cranmer, \emph{{QCD-Aware Recursive Neural
  Networks for Jet Physics}},
  \href{https://doi.org/10.1007/JHEP01(2019)057}{\emph{JHEP} {\bfseries 01}
  (2019) 057} [\href{https://arxiv.org/abs/1702.00748}{{\ttfamily
  1702.00748}}].

\bibitem{Kasieczka:2019dbj}
A.~Butter et~al., \emph{{The Machine Learning landscape of top taggers}},
  \href{https://doi.org/10.21468/SciPostPhys.7.1.014}{\emph{SciPost Phys.}
  {\bfseries 7} (2019) 014} [\href{https://arxiv.org/abs/1902.09914}{{\ttfamily
  1902.09914}}].

\bibitem{Chakraborty:2020yfc}
A.~Chakraborty, S.~H. Lim, M.~M. Nojiri and M.~Takeuchi, \emph{{Neural
  Network-based Top Tagger with Two-Point Energy Correlations and Geometry of
  Soft Emissions}}, \href{https://doi.org/10.1007/JHEP07(2020)111}{\emph{JHEP}
  {\bfseries 07} (2020) 111}
  [\href{https://arxiv.org/abs/2003.11787}{{\ttfamily 2003.11787}}].

\bibitem{Bhattacharya:2020aid}
S.~Bhattacharya, M.~Guchait and A.~H. Vijay, \emph{{Boosted top quark tagging
  and polarization measurement using machine learning}},
  \href{https://doi.org/10.1103/PhysRevD.105.042005}{\emph{Phys. Rev. D}
  {\bfseries 105} (2022) 042005}
  [\href{https://arxiv.org/abs/2010.11778}{{\ttfamily 2010.11778}}].

\bibitem{Ju:2020tbo}
X.~Ju and B.~Nachman, \emph{{Supervised Jet Clustering with Graph Neural
  Networks for Lorentz Boosted Bosons}},
  \href{https://doi.org/10.1103/PhysRevD.102.075014}{\emph{Phys. Rev. D}
  {\bfseries 102} (2020) 075014}
  [\href{https://arxiv.org/abs/2008.06064}{{\ttfamily 2008.06064}}].

\bibitem{Dreyer:2020brq}
F.~A. Dreyer and H.~Qu, \emph{{Jet tagging in the Lund plane with graph
  networks}}, \href{https://doi.org/10.1007/JHEP03(2021)052}{\emph{JHEP}
  {\bfseries 03} (2021) 052}
  [\href{https://arxiv.org/abs/2012.08526}{{\ttfamily 2012.08526}}].

\bibitem{Tannenwald:2020mhq}
B.~Tannenwald, C.~Neu, A.~Li, G.~Buehlmann, A.~Cuddeback, L.~Hatfield et~al.,
  \emph{{Benchmarking Machine Learning Techniques with Di-Higgs Production at
  the LHC}},  \href{https://arxiv.org/abs/2009.06754}{{\ttfamily 2009.06754}}.

\bibitem{Dreyer:2022yom}
F.~A. Dreyer, R.~Grabarczyk and P.~F. Monni, \emph{{Leveraging universality of
  jet taggers through transfer learning}},
  \href{https://doi.org/10.1140/epjc/s10052-022-10469-9}{\emph{Eur. Phys. J. C}
  {\bfseries 82} (2022) 564}
  [\href{https://arxiv.org/abs/2203.06210}{{\ttfamily 2203.06210}}].

\bibitem{Hammad:2022lzo}
A.~Hammad, S.~Khalil and S.~Moretti, \emph{{Search for mono-Higgs signals in
  bb\textasciimacron{} final states using deep neural networks}},
  \href{https://doi.org/10.1103/PhysRevD.107.075027}{\emph{Phys. Rev. D}
  {\bfseries 107} (2023) 075027}
  [\href{https://arxiv.org/abs/2208.10133}{{\ttfamily 2208.10133}}].

\bibitem{Ahmed:2022hct}
I.~Ahmed, A.~Zada, M.~Waqas and M.~U. Ashraf, \emph{{Application of deep
  learning in top pair and single top quark production at the LHC}},
  \href{https://doi.org/10.1140/epjp/s13360-023-04409-z}{\emph{Eur. Phys. J.
  Plus} {\bfseries 138} (2023) 795}
  [\href{https://arxiv.org/abs/2203.12871}{{\ttfamily 2203.12871}}].

\bibitem{Munoz:2022gjq}
J.~M. Munoz, I.~Batatia and C.~Ortner, \emph{{Boost invariant polynomials for
  efficient jet tagging}},
  \href{https://doi.org/10.1088/2632-2153/aca9ca}{\emph{Mach. Learn. Sci.
  Tech.} {\bfseries 3} (2022) 04LT05}
  [\href{https://arxiv.org/abs/2207.08272}{{\ttfamily 2207.08272}}].

\bibitem{He:2023cfc}
M.~He and D.~Wang, \emph{{Quark/gluon discrimination and top tagging with dual
  attention transformer}},
  \href{https://doi.org/10.1140/epjc/s10052-023-12293-1}{\emph{Eur. Phys. J. C}
  {\bfseries 83} (2023) 1116}
  [\href{https://arxiv.org/abs/2307.04723}{{\ttfamily 2307.04723}}].

\bibitem{Aguilar-Saavedra:2023pde}
J.~A. Aguilar-Saavedra, E.~Arganda, F.~R. Joaquim, R.~M. Sand\'a~Seoane and
  J.~F. Seabra, \emph{{Gradient Boosting MUST taggers for highly-boosted
  jets}},  \href{https://arxiv.org/abs/2305.04957}{{\ttfamily 2305.04957}}.

\bibitem{Athanasakos:2023fhq}
D.~Athanasakos, A.~J. Larkoski, J.~Mulligan, M.~P\l{}osko\'n and F.~Ringer,
  \emph{{Is infrared-collinear safe information all you need for jet
  classification?}},  \href{https://arxiv.org/abs/2305.08979}{{\ttfamily
  2305.08979}}.

\bibitem{Grossi:2023fqq}
M.~Grossi, M.~Incudini, M.~Pellen and G.~Pelliccioli, \emph{{Amplitude-assisted
  tagging of longitudinally polarised bosons using wide neural networks}},
  \href{https://doi.org/10.1140/epjc/s10052-023-11931-y}{\emph{Eur. Phys. J. C}
  {\bfseries 83} (2023) 759}
  [\href{https://arxiv.org/abs/2306.07726}{{\ttfamily 2306.07726}}].

\bibitem{Hammad:2024hhm}
A.~Hammad, P.~Ko, C.-T. Lu and M.~Park, \emph{{Exploring Exotic Decays of the
  Higgs Boson to Multi-Photons at the LHC via Multimodal Learning Approaches}},
   \href{https://arxiv.org/abs/2405.18834}{{\ttfamily 2405.18834}}.

\bibitem{Hammad:2024cae}
A.~Hammad and M.~M. Nojiri, \emph{{Streamlined jet tagging network assisted by
  jet prong structure}},
  \href{https://doi.org/10.1007/JHEP06(2024)176}{\emph{JHEP} {\bfseries 06}
  (2024) 176} [\href{https://arxiv.org/abs/2404.14677}{{\ttfamily
  2404.14677}}].

\bibitem{CMS:2017wtu}
{\scshape CMS} collaboration, A.~M. Sirunyan et~al., \emph{{Identification of
  heavy-flavour jets with the CMS detector in pp collisions at 13 TeV}},
  \href{https://doi.org/10.1088/1748-0221/13/05/P05011}{\emph{JINST} {\bfseries
  13} (2018) P05011} [\href{https://arxiv.org/abs/1712.07158}{{\ttfamily
  1712.07158}}].

\bibitem{ATLAS:2017gpy}
{\scshape ATLAS} collaboration, \emph{{Identification of Jets Containing
  $b$-Hadrons with Recurrent Neural Networks at the ATLAS Experiment}}, .

\bibitem{CMS:2020poo}
{\scshape CMS} collaboration, A.~M. Sirunyan et~al., \emph{{Identification of
  heavy, energetic, hadronically decaying particles using machine-learning
  techniques}},
  \href{https://doi.org/10.1088/1748-0221/15/06/P06005}{\emph{JINST} {\bfseries
  15} (2020) P06005} [\href{https://arxiv.org/abs/2004.08262}{{\ttfamily
  2004.08262}}].

\bibitem{ATLAS:2023gog}
{\scshape ATLAS} collaboration, G.~Aad et~al., \emph{{Fast b-tagging at the
  high-level trigger of the ATLAS experiment in LHC Run~3}},
  \href{https://doi.org/10.1088/1748-0221/18/11/P11006}{\emph{JINST} {\bfseries
  18} (2023) P11006} [\href{https://arxiv.org/abs/2306.09738}{{\ttfamily
  2306.09738}}].

\bibitem{Andrews:2021ejw}
M.~Andrews et~al., \emph{{End-to-end jet classification of boosted top quarks
  with the CMS open data}},
  \href{https://doi.org/10.1051/epjconf/202125104030}{\emph{EPJ Web Conf.}
  {\bfseries 251} (2021) 04030}
  [\href{https://arxiv.org/abs/2104.14659}{{\ttfamily 2104.14659}}].

\bibitem{Keicher:2023mer}
P.~Keicher, \emph{{Machine Learning in Top Physics in the ATLAS and CMS
  Collaborations}},  in \emph{{15th International Workshop on Top Quark
  Physics}}, 1, 2023, \href{https://arxiv.org/abs/2301.09534}{{\ttfamily
  2301.09534}}.

\bibitem{Baron:2023yhw}
P.~Baro\v{n}, J.~Kvita, R.~P\v{r}\'\i{}vara, J.~Tome\v{c}ek and R.~Vod\'ak,
  \emph{{Application of Machine Learning Based Top Quark and W Jet Tagging to
  Hadronic Four-Top Final States Induced by SM as well as BSM Processes}},  in
  \emph{{16th International Workshop on Top Quark Physics}}, 10, 2023,
  \href{https://arxiv.org/abs/2310.13009}{{\ttfamily 2310.13009}}.

\bibitem{Hammad:2023sbd}
A.~Hammad, S.~Moretti and M.~Nojiri, \emph{{Multi-scale cross-attention
  transformer encoder for event classification}},
  \href{https://doi.org/10.1007/JHEP03(2024)144}{\emph{JHEP} {\bfseries 03}
  (2024) 144} [\href{https://arxiv.org/abs/2401.00452}{{\ttfamily
  2401.00452}}].

\bibitem{Esmail:2023axd}
W.~Esmail, A.~Hammad and S.~Moretti, \emph{{Sharpening the A \textrightarrow{}
  Z$^{(*)}$h signature of the Type-II 2HDM at the LHC through advanced Machine
  Learning}}, \href{https://doi.org/10.1007/JHEP11(2023)020}{\emph{JHEP}
  {\bfseries 11} (2023) 020}
  [\href{https://arxiv.org/abs/2305.13781}{{\ttfamily 2305.13781}}].

\bibitem{Datta:2019ndh}
K.~Datta, A.~Larkoski and B.~Nachman, \emph{{Automating the Construction of Jet
  Observables with Machine Learning}},
  \href{https://doi.org/10.1103/PhysRevD.100.095016}{\emph{Phys. Rev. D}
  {\bfseries 100} (2019) 095016}
  [\href{https://arxiv.org/abs/1902.07180}{{\ttfamily 1902.07180}}].

\bibitem{Chakraborty:2019imr}
A.~Chakraborty, S.~H. Lim and M.~M. Nojiri, \emph{{Interpretable deep learning
  for two-prong jet classification with jet spectra}},
  \href{https://doi.org/10.1007/JHEP07(2019)135}{\emph{JHEP} {\bfseries 07}
  (2019) 135} [\href{https://arxiv.org/abs/1904.02092}{{\ttfamily
  1904.02092}}].

\bibitem{Kim:2021gtv}
T.~Kim and A.~Martin, \emph{{A $W^\pm$ polarization analyzer from Deep Neural
  Networks}},  \href{https://arxiv.org/abs/2102.05124}{{\ttfamily 2102.05124}}.

\bibitem{Subba:2022czw}
A.~Subba and R.~K. Singh, \emph{{Role of polarizations and spin-spin
  correlations of W's in e-e+\textrightarrow{}W-W+ at s=250\,\,GeV to probe
  anomalous W-W+Z/\ensuremath{\gamma} couplings}},
  \href{https://doi.org/10.1103/PhysRevD.107.073004}{\emph{Phys. Rev. D}
  {\bfseries 107} (2023) 073004}
  [\href{https://arxiv.org/abs/2212.12973}{{\ttfamily 2212.12973}}].

\bibitem{Bogatskiy:2023nnw}
A.~Bogatskiy, T.~Hoffman, D.~W. Miller, J.~T. Offermann and X.~Liu,
  \emph{{Explainable equivariant neural networks for particle physics:
  PELICAN}}, \href{https://doi.org/10.1007/JHEP03(2024)113}{\emph{JHEP}
  {\bfseries 03} (2024) 113}
  [\href{https://arxiv.org/abs/2307.16506}{{\ttfamily 2307.16506}}].

\bibitem{Akar:2020jti}
S.~Akar, T.~J. Boettcher, S.~Carl, H.~F. Schreiner, M.~D. Sokoloff, M.~Stahl
  et~al., \emph{{An updated hybrid deep learning algorithm for identifying and
  locating primary vertices}},
  \href{https://arxiv.org/abs/2007.01023}{{\ttfamily 2007.01023}}.

\bibitem{Shlomi:2020ufi}
J.~Shlomi, S.~Ganguly, E.~Gross, K.~Cranmer, Y.~Lipman, H.~Serviansky et~al.,
  \emph{{Secondary vertex finding in jets with neural networks}},
  \href{https://doi.org/10.1140/epjc/s10052-021-09342-y}{\emph{Eur. Phys. J. C}
  {\bfseries 81} (2021) 540}
  [\href{https://arxiv.org/abs/2008.02831}{{\ttfamily 2008.02831}}].

\bibitem{Goto:2021wmw}
K.~Goto, T.~Suehara, T.~Yoshioka, M.~Kurata, H.~Nagahara, Y.~Nakashima et~al.,
  \emph{{Development of a vertex finding algorithm using Recurrent Neural
  Network}}, \href{https://doi.org/10.1016/j.nima.2022.167836}{\emph{Nucl.
  Instrum. Meth. A} {\bfseries 1047} (2023) 167836}
  [\href{https://arxiv.org/abs/2101.11906}{{\ttfamily 2101.11906}}].

\bibitem{Guiang:2024qzk}
J.~Guiang et~al., \emph{{Improving tracking algorithms with machine learning: a
  case for line-segment tracking at the High Luminosity LHC}},  in
  \emph{{Connecting The Dots 2023}}, 3, 2024,
  \href{https://arxiv.org/abs/2403.13166}{{\ttfamily 2403.13166}}.

\bibitem{Erdmann:2019blf}
J.~Erdmann, \emph{{A tagger for strange jets based on tracking information
  using long short-term memory}},
  \href{https://doi.org/10.1088/1748-0221/15/01/P01021}{\emph{JINST} {\bfseries
  15} (2020) P01021} [\href{https://arxiv.org/abs/1907.07505}{{\ttfamily
  1907.07505}}].

\bibitem{Nakai:2020kuu}
Y.~Nakai, D.~Shih and S.~Thomas, \emph{{Strange Jet Tagging}},
  \href{https://arxiv.org/abs/2003.09517}{{\ttfamily 2003.09517}}.

\bibitem{Erdmann:2020ovh}
J.~Erdmann, O.~Nackenhorst and S.~V. Zei\ss{}ner, \emph{{Maximum performance of
  strange-jet tagging at hadron colliders}},
  \href{https://doi.org/10.1088/1748-0221/16/08/P08039}{\emph{JINST} {\bfseries
  16} (2021) P08039} [\href{https://arxiv.org/abs/2011.10736}{{\ttfamily
  2011.10736}}].

\bibitem{Komiske:2016rsd}
P.~T. Komiske, E.~M. Metodiev and M.~D. Schwartz, \emph{{Deep learning in
  color: towards automated quark/gluon jet discrimination}},
  \href{https://doi.org/10.1007/JHEP01(2017)110}{\emph{JHEP} {\bfseries 01}
  (2017) 110} [\href{https://arxiv.org/abs/1612.01551}{{\ttfamily
  1612.01551}}].

\bibitem{Cheng:2017rdo}
T.~Cheng, \emph{{Recursive Neural Networks in Quark/Gluon Tagging}},
  \href{https://doi.org/10.1007/s41781-018-0007-y}{\emph{Comput. Softw. Big
  Sci.} {\bfseries 2} (2018) 3}
  [\href{https://arxiv.org/abs/1711.02633}{{\ttfamily 1711.02633}}].

\bibitem{Abbas:2020khd}
M.~Abbas, A.~Khan, A.~S. Qureshi and M.~W. Khan, \emph{{Extracting Signals of
  Higgs Boson From Background Noise Using Deep Neural Networks}},
  \href{https://arxiv.org/abs/2010.08201}{{\ttfamily 2010.08201}}.

\bibitem{CMS:2022fxs}
{\scshape CMS} collaboration, A.~Tumasyan et~al., \emph{{Search for Higgs Boson
  and Observation of Z Boson through their Decay into a Charm Quark-Antiquark
  Pair in Boosted Topologies in Proton-Proton Collisions at s=13\,\,TeV}},
  \href{https://doi.org/10.1103/PhysRevLett.131.041801}{\emph{Phys. Rev. Lett.}
  {\bfseries 131} (2023) 041801}
  [\href{https://arxiv.org/abs/2211.14181}{{\ttfamily 2211.14181}}].

\bibitem{Zhang:2023czx}
Z.~Zhang, J.~Liu, J.~Hu, Q.~Wang and U.-G. Mei\ss{}ner, \emph{{Revealing the
  nature of hidden charm pentaquarks with machine learning}},
  \href{https://doi.org/10.1016/j.scib.2023.04.018}{\emph{Sci. Bull.}
  {\bfseries 68} (2023) 981}
  [\href{https://arxiv.org/abs/2301.05364}{{\ttfamily 2301.05364}}].

\bibitem{Goswami:2024xrx}
K.~Goswami, S.~Prasad, N.~Mallick, R.~Sahoo and G.~B. Mohanty, \emph{{A machine
  learning-based study of open-charm hadrons in proton-proton collisions at the
  Large Hadron Collider}},  \href{https://arxiv.org/abs/2404.09839}{{\ttfamily
  2404.09839}}.

\bibitem{Qu:2022mxj}
H.~Qu, C.~Li and S.~Qian, \emph{{Particle Transformer for Jet Tagging}},
  \href{https://arxiv.org/abs/2202.03772}{{\ttfamily 2202.03772}}.

\bibitem{Beauchesne:2023vie}
H.~Beauchesne, Z.-E. Chen and C.-W. Chiang, \emph{{Improving the performance of
  weak supervision searches using transfer and meta-learning}},
  \href{https://doi.org/10.1007/JHEP02(2024)138}{\emph{JHEP} {\bfseries 02}
  (2024) 138} [\href{https://arxiv.org/abs/2312.06152}{{\ttfamily
  2312.06152}}].

\bibitem{Cogan:2014oua}
J.~Cogan, M.~Kagan, E.~Strauss and A.~Schwarztman, \emph{{Jet-Images: Computer
  Vision Inspired Techniques for Jet Tagging}},
  \href{https://doi.org/10.1007/JHEP02(2015)118}{\emph{JHEP} {\bfseries 02}
  (2015) 118} [\href{https://arxiv.org/abs/1407.5675}{{\ttfamily 1407.5675}}].

\bibitem{deOliveira:2015xxd}
L.~de~Oliveira, M.~Kagan, L.~Mackey, B.~Nachman and A.~Schwartzman,
  \emph{{Jet-images \textemdash{} deep learning edition}},
  \href{https://doi.org/10.1007/JHEP07(2016)069}{\emph{JHEP} {\bfseries 07}
  (2016) 069} [\href{https://arxiv.org/abs/1511.05190}{{\ttfamily
  1511.05190}}].

\bibitem{Barnard:2016qma}
J.~Barnard, E.~N. Dawe, M.~J. Dolan and N.~Rajcic, \emph{{Parton Shower
  Uncertainties in Jet Substructure Analyses with Deep Neural Networks}},
  \href{https://doi.org/10.1103/PhysRevD.95.014018}{\emph{Phys. Rev. D}
  {\bfseries 95} (2017) 014018}
  [\href{https://arxiv.org/abs/1609.00607}{{\ttfamily 1609.00607}}].

\bibitem{Komiske:2018oaa}
P.~T. Komiske, E.~M. Metodiev, B.~Nachman and M.~D. Schwartz, \emph{{Learning
  to classify from impure samples with high-dimensional data}},
  \href{https://doi.org/10.1103/PhysRevD.98.011502}{\emph{Phys. Rev. D}
  {\bfseries 98} (2018) 011502}
  [\href{https://arxiv.org/abs/1801.10158}{{\ttfamily 1801.10158}}].

\bibitem{Lee:2019cad}
J.~S.~H. Lee, I.~Park, I.~J. Watson and S.~Yang, \emph{{Quark-Gluon Jet
  Discrimination Using Convolutional Neural Networks}},
  \href{https://doi.org/10.3938/jkps.74.219}{\emph{J. Korean Phys. Soc.}
  {\bfseries 74} (2019) 219}
  [\href{https://arxiv.org/abs/2012.02531}{{\ttfamily 2012.02531}}].

\bibitem{Collado:2020ehf}
J.~Collado, K.~Bauer, E.~Witkowski, T.~Faucett, D.~Whiteson and P.~Baldi,
  \emph{{Learning to isolate muons}},
  \href{https://doi.org/10.1007/JHEP10(2021)200}{\emph{JHEP} {\bfseries 21}
  (2020) 200} [\href{https://arxiv.org/abs/2102.02278}{{\ttfamily
  2102.02278}}].

\bibitem{Li:2020bvf}
J.~Li and H.~Sun, \emph{{An Attention Based Neural Network for Jet Tagging}},
  \href{https://arxiv.org/abs/2009.00170}{{\ttfamily 2009.00170}}.

\bibitem{Li:2020grn}
J.~Li, T.~Li and F.-Z. Xu, \emph{{Reconstructing boosted Higgs jets from event
  image segmentation}},
  \href{https://doi.org/10.1007/JHEP04(2021)156}{\emph{JHEP} {\bfseries 04}
  (2021) 156} [\href{https://arxiv.org/abs/2008.13529}{{\ttfamily
  2008.13529}}].

\bibitem{Filipek:2021qbe}
J.~Filipek, S.-C. Hsu, J.~Kruper, K.~Mohan and B.~Nachman, \emph{{Identifying
  the Quantum Properties of Hadronic Resonances using Machine Learning}},
  \href{https://arxiv.org/abs/2105.04582}{{\ttfamily 2105.04582}}.

\bibitem{Han:2023djl}
T.~Han, I.~M. Lewis, H.~Liu, Z.~Liu and X.~Wang, \emph{{A guide to diagnosing
  colored resonances at hadron colliders}},
  \href{https://doi.org/10.1007/JHEP08(2023)173}{\emph{JHEP} {\bfseries 08}
  (2023) 173} [\href{https://arxiv.org/abs/2306.00079}{{\ttfamily
  2306.00079}}].

\bibitem{Kheddar:2024osf}
H.~Kheddar, Y.~Himeur, A.~Amira and R.~Soualah, \emph{{Image Classification in
  High-Energy Physics: A Comprehensive Survey of Applications to Jet
  Analysis}},  \href{https://arxiv.org/abs/2403.11934}{{\ttfamily 2403.11934}}.

\bibitem{Abdughani:2018wrw}
M.~Abdughani, J.~Ren, L.~Wu and J.~M. Yang, \emph{{Probing stop pair production
  at the LHC with graph neural networks}},
  \href{https://doi.org/10.1007/JHEP08(2019)055}{\emph{JHEP} {\bfseries 08}
  (2019) 055} [\href{https://arxiv.org/abs/1807.09088}{{\ttfamily
  1807.09088}}].

\bibitem{Moreno:2019bmu}
E.~A. Moreno, O.~Cerri, J.~M. Duarte, H.~B. Newman, T.~Q. Nguyen, A.~Periwal
  et~al., \emph{{JEDI-net: a jet identification algorithm based on interaction
  networks}}, \href{https://doi.org/10.1140/epjc/s10052-020-7608-4}{\emph{Eur.
  Phys. J. C} {\bfseries 80} (2020) 58}
  [\href{https://arxiv.org/abs/1908.05318}{{\ttfamily 1908.05318}}].

\bibitem{Bernreuther:2020vhm}
E.~Bernreuther, T.~Finke, F.~Kahlhoefer, M.~Kr\"amer and A.~M\"uck,
  \emph{{Casting a graph net to catch dark showers}},
  \href{https://doi.org/10.21468/SciPostPhys.10.2.046}{\emph{SciPost Phys.}
  {\bfseries 10} (2021) 046}
  [\href{https://arxiv.org/abs/2006.08639}{{\ttfamily 2006.08639}}].

\bibitem{Iiyama:2020wap}
Y.~Iiyama et~al., \emph{{Distance-Weighted Graph Neural Networks on FPGAs for
  Real-Time Particle Reconstruction in High Energy Physics}},
  \href{https://doi.org/10.3389/fdata.2020.598927}{\emph{Front. Big Data}
  {\bfseries 3} (2020) 598927}
  [\href{https://arxiv.org/abs/2008.03601}{{\ttfamily 2008.03601}}].

\bibitem{Heintz:2020soy}
A.~Heintz et~al., \emph{{Accelerated Charged Particle Tracking with Graph
  Neural Networks on FPGAs}},  in \emph{{34th Conference on Neural Information
  Processing Systems}}, 11, 2020,
  \href{https://arxiv.org/abs/2012.01563}{{\ttfamily 2012.01563}}.

\bibitem{ExaTrkX:2020nyf}
{\scshape Exa.TrkX} collaboration, X.~Ju et~al., \emph{{Graph Neural Networks
  for Particle Reconstruction in High Energy Physics detectors}},  in
  \emph{{33rd Annual Conference on Neural Information Processing Systems}}, 3,
  2020, \href{https://arxiv.org/abs/2003.11603}{{\ttfamily 2003.11603}}.

\bibitem{Pata:2021oez}
J.~Pata, J.~Duarte, J.-R. Vlimant, M.~Pierini and M.~Spiropulu, \emph{{MLPF:
  Efficient machine-learned particle-flow reconstruction using graph neural
  networks}}, \href{https://doi.org/10.1140/epjc/s10052-021-09158-w}{\emph{Eur.
  Phys. J. C} {\bfseries 81} (2021) 381}
  [\href{https://arxiv.org/abs/2101.08578}{{\ttfamily 2101.08578}}].

\bibitem{Verma:2021ceh}
Y.~Verma and S.~Jena, \emph{{Jet characterization in Heavy Ion Collisions by
  QCD-Aware Graph Neural Networks}},
  \href{https://arxiv.org/abs/2103.14906}{{\ttfamily 2103.14906}}.

\bibitem{Atkinson:2021jnj}
O.~Atkinson, A.~Bhardwaj, S.~Brown, C.~Englert, D.~J. Miller and P.~Stylianou,
  \emph{{Improved constraints on effective top quark interactions using edge
  convolution networks}},
  \href{https://doi.org/10.1007/JHEP04(2022)137}{\emph{JHEP} {\bfseries 04}
  (2022) 137} [\href{https://arxiv.org/abs/2111.01838}{{\ttfamily
  2111.01838}}].

\bibitem{Gong:2022lye}
S.~Gong, Q.~Meng, J.~Zhang, H.~Qu, C.~Li, S.~Qian et~al., \emph{{An efficient
  Lorentz equivariant graph neural network for jet tagging}},
  \href{https://doi.org/10.1007/JHEP07(2022)030}{\emph{JHEP} {\bfseries 07}
  (2022) 030} [\href{https://arxiv.org/abs/2201.08187}{{\ttfamily
  2201.08187}}].

\bibitem{Ma:2022bvt}
F.~Ma, F.~Liu and W.~Li, \emph{{Jet tagging algorithm of graph network with
  Haar pooling message passing}},
  \href{https://doi.org/10.1103/PhysRevD.108.072007}{\emph{Phys. Rev. D}
  {\bfseries 108} (2023) 072007}
  [\href{https://arxiv.org/abs/2210.13869}{{\ttfamily 2210.13869}}].

\bibitem{Bogatskiy:2022czk}
A.~Bogatskiy, T.~Hoffman, D.~W. Miller and J.~T. Offermann, \emph{{PELICAN:
  Permutation Equivariant and Lorentz Invariant or Covariant Aggregator Network
  for Particle Physics}},  \href{https://arxiv.org/abs/2211.00454}{{\ttfamily
  2211.00454}}.

\bibitem{Builtjes:2022usj}
L.~Builtjes, S.~Caron, P.~Moskvitina, C.~Nellist, R.~R. de~Austri, R.~Verheyen
  et~al., \emph{{Attention to the strengths of physical interactions:
  Transformer and graph-based event classification for particle physics
  experiments}},  \href{https://arxiv.org/abs/2211.05143}{{\ttfamily
  2211.05143}}.

\bibitem{DiBello:2022iwf}
F.~A. Di~Bello et~al., \emph{{Reconstructing particles in jets using set
  transformer and hypergraph prediction networks}},
  \href{https://doi.org/10.1140/epjc/s10052-023-11677-7}{\emph{Eur. Phys. J. C}
  {\bfseries 83} (2023) 596}
  [\href{https://arxiv.org/abs/2212.01328}{{\ttfamily 2212.01328}}].

\bibitem{Mokhtar:2022pwm}
F.~Mokhtar, R.~Kansal and J.~Duarte, \emph{{Do graph neural networks learn
  traditional jet substructure?}},  in \emph{{36th Conference on Neural
  Information Processing Systems}: {Workshop on Machine Learning and the
  Physical Sciences}}, 11, 2022,
  \href{https://arxiv.org/abs/2211.09912}{{\ttfamily 2211.09912}}.

\bibitem{Huang:2023ssr}
A.~Huang, X.~Ju, J.~Lyons, D.~Murnane, M.~Pettee and L.~Reed,
  \emph{{Heterogeneous Graph Neural Network for identifying hadronically
  decayed tau leptons at the High Luminosity LHC}},
  \href{https://doi.org/10.1088/1748-0221/18/07/P07001}{\emph{JINST} {\bfseries
  18} (2023) P07001} [\href{https://arxiv.org/abs/2301.00501}{{\ttfamily
  2301.00501}}].

\bibitem{Duperrin:2023elp}
{\scshape ATLAS} collaboration, A.~Duperrin, \emph{{Flavour tagging with graph
  neural networks with the ATLAS detector}},  in \emph{{30th International
  Workshop on Deep-Inelastic Scattering and Related Subjects}}, 6, 2023,
  \href{https://arxiv.org/abs/2306.04415}{{\ttfamily 2306.04415}}.

\bibitem{Konar:2023ptv}
P.~Konar, V.~S. Ngairangbam and M.~Spannowsky, \emph{{Hypergraphs in LHC
  phenomenology \textemdash{} the next frontier of IRC-safe feature
  extraction}}, \href{https://doi.org/10.1007/JHEP01(2024)113}{\emph{JHEP}
  {\bfseries 01} (2024) 113}
  [\href{https://arxiv.org/abs/2309.17351}{{\ttfamily 2309.17351}}].

\bibitem{Qasim:2019otl}
S.~R. Qasim, J.~Kieseler, Y.~Iiyama and M.~Pierini, \emph{{Learning
  representations of irregular particle-detector geometry with
  distance-weighted graph networks}},
  \href{https://doi.org/10.1140/epjc/s10052-019-7113-9}{\emph{Eur. Phys. J. C}
  {\bfseries 79} (2019) 608}
  [\href{https://arxiv.org/abs/1902.07987}{{\ttfamily 1902.07987}}].

\bibitem{qi2017pointnet}
C.~R. Qi, H.~Su, K.~Mo and L.~J. Guibas, \emph{Pointnet: Deep learning on point
  sets for 3d classification and segmentation},  in \emph{Proceedings of the
  IEEE conference on computer vision and pattern recognition}, pp.~652--660,
  2017.

\bibitem{zaheer2017deep}
M.~Zaheer, S.~Kottur, S.~Ravanbakhsh, B.~Poczos, R.~R. Salakhutdinov and A.~J.
  Smola, \emph{Deep sets}, {\emph{Advances in neural information processing
  systems} {\bfseries 30} (2017) }.

\bibitem{Komiske:2018cqr}
P.~T. Komiske, E.~M. Metodiev and J.~Thaler, \emph{{Energy Flow Networks: Deep
  Sets for Particle Jets}},
  \href{https://doi.org/10.1007/JHEP01(2019)121}{\emph{JHEP} {\bfseries 01}
  (2019) 121} [\href{https://arxiv.org/abs/1810.05165}{{\ttfamily
  1810.05165}}].

\bibitem{zhang2022set}
L.~Zhang, V.~Tozzo, J.~Higgins and R.~Ranganath, \emph{Set norm and equivariant
  skip connections: Putting the deep in deep sets},  in \emph{International
  Conference on Machine Learning}, pp.~26559--26574, PMLR, 2022.

\bibitem{Qu:2019gqs}
H.~Qu and L.~Gouskos, \emph{{ParticleNet: Jet Tagging via Particle Clouds}},
  \href{https://doi.org/10.1103/PhysRevD.101.056019}{\emph{Phys. Rev. D}
  {\bfseries 101} (2020) 056019}
  [\href{https://arxiv.org/abs/1902.08570}{{\ttfamily 1902.08570}}].

\bibitem{Shmakov:2021qdz}
A.~Shmakov, M.~J. Fenton, T.-W. Ho, S.-C. Hsu, D.~Whiteson and P.~Baldi,
  \emph{{SPANet: Generalized permutationless set assignment for particle
  physics using symmetry preserving attention}},
  \href{https://doi.org/10.21468/SciPostPhys.12.5.178}{\emph{SciPost Phys.}
  {\bfseries 12} (2022) 178}
  [\href{https://arxiv.org/abs/2106.03898}{{\ttfamily 2106.03898}}].

\bibitem{Finke:2023veq}
T.~Finke, M.~Kr\"amer, A.~M\"uck and J.~T\"onshoff, \emph{{Learning the
  language of QCD jets with transformers}},
  \href{https://doi.org/10.1007/JHEP06(2023)184}{\emph{JHEP} {\bfseries 06}
  (2023) 184} [\href{https://arxiv.org/abs/2303.07364}{{\ttfamily
  2303.07364}}].

\bibitem{Mikuni:2021pou}
V.~Mikuni and F.~Canelli, \emph{{Point cloud transformers applied to collider
  physics}}, \href{https://doi.org/10.1088/2632-2153/ac07f6}{\emph{Mach. Learn.
  Sci. Tech.} {\bfseries 2} (2021) 035027}
  [\href{https://arxiv.org/abs/2102.05073}{{\ttfamily 2102.05073}}].

\bibitem{Spinner:2024hjm}
J.~Spinner, V.~Bres\'o, P.~de~Haan, T.~Plehn, J.~Thaler and J.~Brehmer,
  \emph{{Lorentz-Equivariant Geometric Algebra Transformers for High-Energy
  Physics}},  \href{https://arxiv.org/abs/2405.14806}{{\ttfamily 2405.14806}}.

\bibitem{Bhardwaj:2024djv}
A.~Bhardwaj, C.~Englert, W.~Naskar, V.~S. Ngairangbam and M.~Spannowsky,
  \emph{{Equivariant, safe and sensitive \textemdash{} graph networks for new
  physics}}, \href{https://doi.org/10.1007/JHEP07(2024)245}{\emph{JHEP}
  {\bfseries 07} (2024) 245}
  [\href{https://arxiv.org/abs/2402.12449}{{\ttfamily 2402.12449}}].

\bibitem{vanBeekveld:2024wws}
M.~van Beekveld et~al., \emph{{A new standard for the logarithmic accuracy of
  parton showers}},  \href{https://arxiv.org/abs/2406.02661}{{\ttfamily
  2406.02661}}.

\bibitem{Bierlich:2022pfr}
C.~Bierlich et~al., \emph{{A comprehensive guide to the physics and usage of
  PYTHIA 8.3}},
  \href{https://doi.org/10.21468/SciPostPhysCodeb.8}{\emph{SciPost Phys.
  Codeb.} {\bfseries 2022} (2022) 8}
  [\href{https://arxiv.org/abs/2203.11601}{{\ttfamily 2203.11601}}].

\bibitem{deFavereau:2013fsa}
{\scshape DELPHES 3} collaboration, J.~de~Favereau, C.~Delaere, P.~Demin,
  A.~Giammanco, V.~Lema\^\i{}tre, A.~Mertens et~al., \emph{{DELPHES 3, A
  modular framework for fast simulation of a generic collider experiment}},
  \href{https://doi.org/10.1007/JHEP02(2014)057}{\emph{JHEP} {\bfseries 02}
  (2014) 057} [\href{https://arxiv.org/abs/1307.6346}{{\ttfamily 1307.6346}}].

\bibitem{kornblith2019similarity}
S.~Kornblith, M.~Norouzi, H.~Lee and G.~Hinton, \emph{Similarity of neural
  network representations revisited},  in \emph{International conference on
  machine learning}, pp.~3519--3529, PMLR, 2019.

\bibitem{selvaraju2020grad}
R.~R. Selvaraju, M.~Cogswell, A.~Das, R.~Vedantam, D.~Parikh and D.~Batra,
  \emph{Grad-cam: visual explanations from deep networks via gradient-based
  localization}, {\emph{International journal of computer vision} {\bfseries
  128} (2020) 336}.

\bibitem{zhou2016learning}
B.~Zhou, A.~Khosla, A.~Lapedriza, A.~Oliva and A.~Torralba, \emph{Learning deep
  features for discriminative localization},  in \emph{Proceedings of the IEEE
  conference on computer vision and pattern recognition}, pp.~2921--2929, 2016.

\end{thebibliography}\endgroup

\end{document}